\documentclass[twocolumn]{emulateapj}
\usepackage{natbib}
\usepackage{amsmath}
\usepackage{epstopdf}
\usepackage{color}
\usepackage{soul}

\newcommand{\ang}{\mbox{\AA}} 

\begin{document}
\title{H$\alpha$ Absorption in Transiting Exoplanet Atmospheres}
\author{Duncan Christie, Phil Arras, and Zhi-Yun Li}
\affil{Department of Astronomy, University of Virginia}
\affil{P.O. Box 400325 \\ Charlottesville, VA \\ 22904-4325}
\email{dac5zm@virginia.edu, pla7y@virginia.edu, zl4h@virginia.edu}

\begin{abstract}
Absorption of stellar H$\alpha$ by the upper atmosphere of the planet HD189733b 
has recently been detected by Jensen et al. 
Motivated by this observation, we have developed a model for atomic hydrogen in the $n=2$ state and compared the resulting
H$\alpha$ line profile to the observations.
The model atmosphere is in hydrostatic balance, as well as thermal and photoionization equilibrium. 
Collisional and radiative transitions are included in the determination of the $n=2$ state level population. 
We find that H$\alpha$ absorption is dominated by an optical depth $\tau \sim 1$ shell, composed of hydrogen in the metastable 2s state that is located below the hydrogen ionization layer. 
The number density of the 2s state within the shell is found to vary slowly with radius, while that of the 1s state falls rapidly. Thus while the Ly$\alpha$ absorption, for a certain wavelength, occurs inside a relatively well defined impact parameter, the contribution to H$\alpha$ absorption is roughly uniform over the entire atomic hydrogen layer. The model can approximately reproduce the observed Ly$\alpha$ and H$\alpha$ integrated transit depths for HD189733b by using an ionization rate enhanced over that expected for the star by an order of magnitude. For HD 209458b, we are unable to explain the asymmetric H$\alpha$ line profile observed by Jensen et al., as the model produces a symmetric line profile with transit depth comparable to that of HD 189733b. In an appendix, we study the effect of the stellar Ly$\alpha$ absorption on the net cooling rate.

\end{abstract}

\section{Introduction}

Absorption\footnote{ We will use the term ``absorption" thoughout the paper, even though at the low densities of interest each $n=2 \rightarrow 3$ radiative transition is likely followed by a $n=3 \rightarrow 2$ (resonant scattering) or $3 \rightarrow 1$ (resonance fluorescence) radiative transition. Either outcome will prevent the stellar H$\alpha$ photon from reaching the observer.}
of  starlight by a transiting planet provides a probe of the atmospheric composition and structure. The Ly$\alpha$ and H$\alpha$ transitions of hydrogen may potentially give rise to large transit depths,  as the combination of large cross section and hydrogen abundance put the optical depth unity surfaces at much higher altitudes ($\mu \rm bar - nbar$ pressure levels) than the continuum photosphere (\rm mbar - bar). 

The detection of Ly$\alpha$ absorption \citep{2003Natur.422..143V, 2004ApJ...604L..69V,2008ApJ...676L..57V,2007ApJ...671L..61B,2008ApJ...688.1352B,2008A&A...483..933E,2010A&A...514A..72L,2012A&A...547A..18E,2012A&A...543L...4L} due to atoms in the ground state traces the bulk of the atomic population, and hence the density profile of the planet's atmosphere. By contrast, the recent detection of H$\alpha$ absorption by \citet{2012ApJ...751...86J} reveals the small population in the $n=2$ excited state. As the number density in the $n=2$ state depends sensitively on the excitation rates due to collisions and radiative pumping by the stellar photons, it is a much more sensitive probe of the atmosphere, in particular the temperature profile. Hence, the Ly$\alpha$ and H$\alpha$ lines are complementary probes of the atomic hydrogen layer. 
 
The goal of this paper is to construct a detailed model of the $n=2$ state population by which to understand the \citet{2012ApJ...751...86J} H$\alpha$ transit depth  of HD 189733b. Section \ref{sec:review} reviews the relevant observations of Ly$\alpha$ and H$\alpha$ observations for HD 209458b and HD 189733b. Section \ref{sec:model} presents the details of the atmospheric model. Section \ref{sec:results} presents numerical results for HD 189733b, and a detailed comparison to the \citet{2012ApJ...751...86J} data. Section \ref{sec:hd209} presents the predictions of our model for HD 209458b, and they disagree significantly with the data.  Section \ref{sec:balmercont} addresses the possibility of observing Balmer continuum absorption. We summarize our results and compare to previous investigations in Section \ref{sec:summary}. In the appendix, we study the effect of the stellar Ly$\alpha$ on the cooling rate.

 \section{ Review of the Observations }
 \label{sec:review} 
 
 We first review the observations of H$\alpha$ absorption that motivate our work, as well as previous attempts to determine excited state abundance, or excitation temperature, from the observations. These previous estimates of density and temperature will be compared to the calculations in this paper in Section \ref{sec:summary}.\
 
 
The initial detection of Ly$\alpha$ absorption during the transit of HD209458b \citep{2003Natur.422..143V} indicated a $15\%$ decrease in flux over a $5.1{\rm\, \ang}$  wavelength band around line center.  The implied occultation radius of $4.3 {\rm \,R_J}$, where $R_J=7.149\times 10^9\,{\rm cm}$ is the radius of Jupiter, is larger than the Roche lobe radius of $3.6 {\rm \,R_J}$, leading the authors to suggest that the planet must be in a state of Roche lobe overflow.  Due to absorption by the interstellar medium  and geocoronal emission, the center of the Ly$\alpha$ line cannot be used, and absorption is only reliably observed outside $\sim 75 \, {\rm km/s}$ from line center, indicating either a population of ``hot hydrogen" ($T \sim 10^6\,{\rm K}$) formed through charge exchange with stellar wind protons \citep{2008Natur.451..970H,2010ApJ...709..670E, 2012arXiv1206.5003T} or large columns of ``warm hydrogen" ($T \sim 10^4\,{\rm K}$;  \citealt{2004Icar..170..167Y}),  which yield a saturated line. This paper attempts to explain H$\alpha$ transit spectra by the latter scenario of a large column of warm hydrogen.



For $T \sim 10^4\ {\rm K}$, the observed Ly$\alpha$ absorption is well out on the damping wing of the line, and hence the cross section is not a strong function of temperature. The Ly$\alpha$ transit depth is, however, dependent on the temperature through the atmospheric scale height. As we will show, the H$\alpha$ absorption is expected to have a stronger dependence on temperature due to the collisional excitation rate of the $n=1$ to $n=2$ transition.

\begin{deluxetable}{lc}
\tablecolumns{2}
\tablewidth{0pc}
\tablecaption{HD 189733b Parameters$^a$}
\startdata
Mass ($M_{\rm J}$) & 1.138 \\
Radius ($R_{\rm J}$) & 1.138 \\
a (AU) & 0.031 \\
\enddata
\tablenotetext{a}{Source: exoplanet.eu}
\label{table:189733b}
\end{deluxetable}

The first attempt to detect H$\alpha$ was by \citet{2004PASJ...56..655W} who observed HD209458. For the remainder of this section, we summarize their observational results, assumptions,  and inferences.   Integrating over a $5.1\ang$ band, they find an equivalent width

\begin{equation}
W_\lambda \equiv \int d\lambda \left( \frac{F_{\rm out}(\lambda)-F_{\rm in}(\lambda)}{F_{\rm out}(\lambda)} \right) < 1.7 {\rm \, m\ang},
\label{Eqn:width}
\end{equation}

\noindent where $F_{\rm in}$ and $F_{\rm out}$ are the flux in and out of transit, respectively.  If the $n=2$ state hydrogen occults an area $\Delta A$, and the area of the star is $A$, then in the optically thin limit, eq. (\ref{Eqn:width}) can be used to relate the column density of the $n=2$ state, $N_2$, to the equivalent width as
\footnote{ We believe eq.6 of \citet{2004PASJ...56..655W} and eq.4 in \citet{2012ApJ...751...86J}
should have $\Delta A/A$ on the right hand side, not the left, as
the drop in observed flux is proportional to the
occulting area, not its inverse. Their estimates of
$N_2$ should be multiplied by the factor
$(A/\Delta A)^2$, implying columns larger by a factor $10-100$. }
\begin{equation}
W_{\rm H_\alpha} = \frac{\pi e^2}{m_ec^2}f_{23}\lambda_{\rm H_\alpha}^2N_2 \frac{\Delta A}{A} \, .
\label{Eqn:WidthColumnDens}
\end{equation}

\noindent Here $\lambda_{\rm H_\alpha}=6562.8\ \ang$ is the wavelength of the $\rm H\alpha$ transition, and $f_{\rm 23}=0.64$ is the oscillator strength.
Using a fractional area $\Delta A/A=0.15$
from Ly$\alpha$ measurements \citep{2003Natur.422..143V}, and
assuming the same occulting area for Ly$\alpha$ and H$\alpha$ lines,
the column density can be constrained to be $N_2 \la 2.4 \times 10^{11}\, \rm cm^{-2}$.

H$\alpha$ absorption was subsequently detected by \citet{2012ApJ...751...86J} in the transits of HD 209458b and HD 189733b.
The line profile of  HD 209458b showed an excess to the blue and a deficit to the red of line center, for which we have no simple explanation. The line profile of HD 189733b offered no such consternation, exhibiting a symmetric absorption feature about line center.  
HD 189733b is the focus of our modeling effort.

To quantify the absorption, \citet{2012ApJ...751...86J} introduce the absorption measure $M_{\rm abs}$, defined as

\begin{equation}
M_{\rm abs} = \left< S_T\right>_{\rm central}-\frac{\left<S_T\right>_{\rm blue} + \left<S_T\right>_{\rm red}}{2}
\end{equation}

\noindent where

\begin{equation}
\left< S_T \right>_{\rm i} = \left(\frac{F_{\rm in}}{F_{\rm out}}\right)_{\rm i} - 1
\end{equation}

\noindent and i={\em blue}, {\em red}, {\em central} indicates the domain of the wavelength integration.  The purpose of subtracting off the bands just outside the line center is to derive the drop in flux due to the upper atmosphere, where the H$\alpha$ absorption takes place;
absorption of the neighboring continuum takes place much deeper
in the planet's atmosphere, near the continuum photospheres at mbar-bar pressures. In their analysis, \citet{2012ApJ...751...86J} considered three $16 {\rm \, \ang}$ bands with the {\em central} band covering the H$\alpha$ line and {\em blue} and {\em red} being adjacent bands at shorter and longer wavelengths, respectively. The absorption measure $M_{\rm abs}$ can be related back to  equivalent width by $W_\lambda \approx M_{\rm abs}\Delta \lambda$. For HD 189733b, the absorption measure is $M_{\rm abs}=(-8.72\pm 1.48)\times 10^{-4}$ \citep{2012ApJ...751...86J}.

 \citet{2004PASJ...56..655W}  and \citet{2012ApJ...751...86J} estimate the excitation temperature, $T_{\rm exc}$, starting with the Boltzmann distribution
for the number densities $n_1$ and $n_2$ of atoms in states $n=1$ and $n=2$:
\begin{eqnarray}
\frac{n_2}{n_1} & =& \frac{g_2}{g_1} e^{-10.2\ {\rm eV}/k_B T_{\rm exc}},
\label{Eqn:Boltz}
\end{eqnarray}
where $g_2=8$ and $g_1=2$ are the degeneracies of each state, including both 2s and 2p upper states, and the energy difference is $10.2$eV.
The authors then estimate the left hand side by equating $n_2/n_1 \simeq N_2/N_1$, where $N_1$ is the column density of ground state hydrogen.
This assumption relies on the distribution of $n_1$ and $n_2$ being similar in the atmosphere, so that the effective path length in the integration is comparable. Lastly, they assume that $N_2/N_1 \simeq  W_{\rm H\alpha}/ W_{\rm Ly\alpha} = 0.0128\ang/ 0.32\ang = 0.04$, 
which is valid if both  transitions are optically thin and the occulting area $\Delta A$ is the same for each.
Plugging this into the left hand side and solving then gives $T_{\rm exc}=2.6\times 10^4\,{\rm K}$. This high excitation temperature is not achieved for thermal gas in any published models to date (e.g. \citealt{2004Icar..170..167Y,2009ApJ...693...23M,2007P&SS...55.1426G}). We will revisit the estimates of $T_{\rm exc}$ in section \ref{sec:summary} in the context of our model for the atmospheric structure and level populations.

\section{Model for the Atmosphere}
\label{sec:model}

In this section we present our model for the density and temperature profiles and the hydrogen level populations. Section \ref{sec:basic} 
outlines the different layers of the atmosphere of interest. Section \ref{sec:iontherm} discusses thermal and ionization equilibrium. Section \ref{sec:levelpop} presents the rate-equilibrium equations which determine the ground and excited state populations of hydrogen. Section \ref{sec:numerics} discusses the numerical solution of the coupled equations of hydrostatic balance, thermal equilibrium photoionization equilibrium and detailed balance for the level populations.

\subsection{Basic Structure of the Atmosphere}
\label{sec:basic}

\begin{figure}
\epsscale{1.0}
\plotone{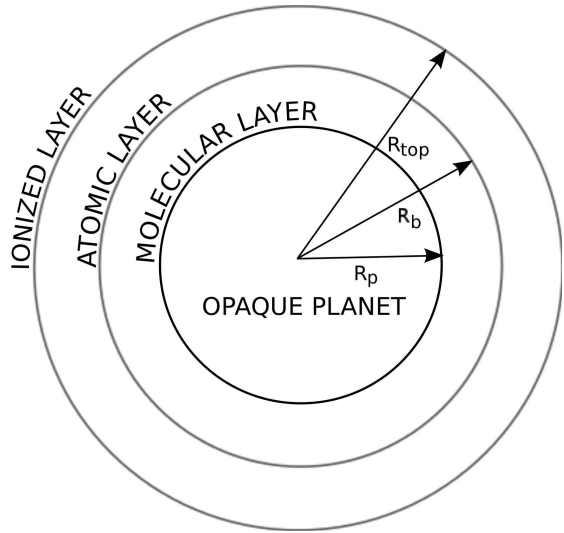}
\caption{ Cartoon showing the different layers in the atmosphere.}
\label{Fig:cartoon}
\end{figure}

We consider a spherically symmetric atmosphere
above the photospheric radius at $r=R_p$, where $r$ is the
spherical radius. The atmosphere is divided
into three zones (see Figure~\ref{Fig:cartoon}), according to
the state of hydrogen: molecular, atomic and
ionized. Inside the
photospheric radius, $r < R_{\rm p}$, we consider the planet to be
opaque to H$\alpha$.  The shell $R_{\rm p} < r < R_{\rm b}$ is composed
mainly of molecular gas, due to the low temperature. Assuming a Boltzmann
distribution and a mean temperature of $T_{\rm eq} = 1200\,{\rm K}$, the
equilibrium temperature (for zero albedo) of HD189733b, the relative abundance of $n=2$
hydrogen is $n_{\rm 2}/n_{\rm 1}=5.65\times 10^{-43}$, where $n_{\rm 1}$ and $n_{\rm 2}$ are
the number density of hydrogen in the $n=1$ and $n=2$ states, respectively (we use $n$
for both number density and radial quantum number. The meaning is clear from
the context).    
Due to the small abundances and low temperatures, we assume that the contribution of the molecular
layer to the H$\alpha$ absorption is negligible. 
The transition from molecular to atomic hydrogen takes place at a
radius $R_b$, and it is this ``atomic" layer at $r > R_b$ and $P \la
1\ {\rm \mu bar}$ that gives rise to the H$\alpha$ absorption 
in our model. At pressures $P \la 1\ \rm nbar$ hydrogen will be 
mostly photoionized, forming an ``ionized'' layer that contributes 
little to the H$\alpha$ absorption. 

A parameter that enters our model is the base radius of the atomic
layer $R_b$. It is determined by the photospheric radius $R_p$ and
the thickness of the molecular layer $R_b-R_p$. If the molecular
layer is isothermal at temperature $T $, the thickness is
\begin{equation}
R_{\rm b}-R_{\rm p} \simeq \frac{k_{\rm B}T R_{\rm p}^2}{\mu m_{\rm H}G M_{\rm p} }\ln\left(\frac{P_{\rm p}}{P_{\rm b}}\right),
\label{Eqn:RadIncrease}
\end{equation}
where $\mu \simeq 2.3$ is the mean molecular weight, $M_p$ is the mass of the planet, $P_{\rm p}$
is the pressure of the optical photosphere, and $P_{\rm b}$ is the pressure at the molecular-to-atomic
transition. For $T=T_{\rm eq}$, $P_{\rm b}=1\ {\rm \mu bar}$ and $P_{\rm p}= 1\ \rm bar$,
the thickness is $R_{\rm b} - R_{\rm p} = 0.032\,R_{\rm p}$.
An upper bound to the thickness is found using the
dissociation temperature at which the gas transitions from molecular to atomic. Plugging
$P_{\rm b} = 1\, {\rm \mu bar}$ into the Saha equation, and using the roto-vibrational
energies from \citet{1989ApJ...336..495B},  we find $T = 1934\,{\rm K}$
and $R_{\rm b}-R_{\rm p}=0.054\,R_{\rm p}$. Since the equilibrium and dissociation temperature estimates
differ by only 40\%, and both give altitudes which are a small fraction of $R_{\rm p}$, we conclude the thickness of the molecular
layer will not greatly affect the H$\alpha$ transit depth. Henceforth we use  $R_{\rm b} - R_{\rm p} = 0.054\,R_{\rm p}$ in 
numerical calculations for HD 189733b.

\subsection{ Ionization and Thermal Equilibrium }
\label{sec:iontherm}

\begin{deluxetable*}{lcccl}
 \tabletypesize{\tiny}%
\tablecolumns{5}
\tablewidth{0pc}
\tablecaption{Relevant Reactions}
\tablehead{\colhead{Number} & \colhead{Reaction} & \colhead{Symbol} & \colhead{Rate} & \colhead{Reference} }
\startdata
R1 & ${\rm H}_{\rm 1s} + \gamma \rightarrow {\rm e}^- + {\rm H}^+$ & $\Gamma_{\rm 1s}$ & See Eq. \ref{Eqn:Ionize} &  \\
R2 & $e^-  + p  \rightarrow {\rm H} + \gamma$ & $\alpha_{\rm B}$  & $2.54\times 10^{-13} (T/10^4 {\rm \, K})^{-0.8164-0.0208\log(T/10^4\,{\rm K})} \,{\rm cm^3\, s^{-1}}$ &  \citet{2011piim.book.....D} \\
R3 & ${\rm H}_{\rm 1s} + e^-  \rightarrow {\rm H}_{\rm 2s} + {\rm e}^-$ & $C_{\rm 1s\rightarrow 2s}$ & $1.21\times 10^{-8}\left(10^4\,{\rm K}/T\right)^{0.455}e^{-118400/T}\,{\rm cm^3\, s^{-1}}$ & \citet{janev2003collision} \\
R4 & ${\rm H}_{\rm 1s} + e^- \rightarrow {\rm H}_{\rm 2p} + {\rm e}^-$ & $C_{\rm 1s\rightarrow 2p}$ & $1.71\times 10^{-8}\left(10^4\,{\rm K}/T\right)^{0.077}e^{-118400/T}\,{\rm cm^3\, s^{-1}}$ & \citet{janev2003collision} \\
R5 & ${\rm H}_{\rm 2s} + e^- \rightarrow {\rm H}_{\rm 2p} + {\rm e}^{-}$ & $C_{\rm 2s \rightarrow 2p} $ & $6.21\times 10^{-5}\left(\log\left(T/T_0\right)-\gamma\right)/\sqrt{T}\,{\rm cm^3\, s^{-1}}$ & \citet{janev2003collision}\\
R6 & ${\rm H}_{\rm 2s} + \gamma \rightarrow {\rm e}^- + {\rm H}^+ $ & $\Gamma_{\rm 2s}$ & See Eq. \ref{Eqn:Ionize2s2p} &  \\
R7 & ${\rm H}_{\rm 2p} + \gamma \rightarrow {\rm e}^- + {\rm H}^+ $ & $\Gamma_{\rm 2p}$ & See Eq. \ref{Eqn:Ionize2s2p} & \\
R8 & $e^{-} + {\rm H}^+ \rightarrow {\rm H}_{\rm 2s}+\gamma$ & $\alpha_{\rm 2s}$ & $\left(0.282+0.047(T/10^4\,{\rm K})-0.006(T/10^4\,{\rm K})^2\right)\alpha_{\rm B}$ & \citet{2011piim.book.....D} \\ 
R9 & $e^{-} + {\rm H}^+ \rightarrow {\rm H}_{\rm 2p}+\gamma$ & $\alpha_{\rm 2p}$ & $\alpha_{\rm B}-\alpha_{\rm 2s}$ & \citet{2011piim.book.....D}\\ 
R10 & ${\rm H}_{\rm 2s} \rightarrow {\rm H}_{\rm 1s} + 2\gamma$ & $A_{\rm 2s\rightarrow 1s}$ & $8.26\,\,{\rm s^{-1}}$ &  \citet{2006agna.book.....O} \\
R11 & ${\rm H}_{\rm 2p} \rightarrow {\rm H}_{\rm 1s} + \gamma$ & $A_{\rm 2p\rightarrow 1s}$ & $6.3\times 10^8\,\,{\rm s^{-1}}$ &  \citet{2006agna.book.....O} \\
R12 & Ly$\alpha$ cooling & $\Lambda_{\rm Ly\alpha}(T)$ & $10.2\,{\rm eV}\left(C_{\rm 1s\rightarrow 2s}+C_{\rm 1s\rightarrow 2p}\right)$ & \\
\enddata
\label{Tbl:Reacs}
\end{deluxetable*}

The electron and ion abundances are set by rate equilibrium between photoionization of ground-state hydrogen and radiative recombination,

\begin{equation}
\Gamma_{\rm 1s}\left(N_{\rm 1s}\right)n_{\rm 1s} = \alpha_{\rm B}(T) n_{\rm e}^2\,\,.
\label{Eqn:IonizeRecomb}
\end{equation} 

Here $\Gamma_{\rm 1s}$ is the photoionization rate from the 1s state, defined in eq. \ref{Eqn:Ionize} below, and 
$\alpha_{\rm B}$ is the Case B recombination coefficient defined in Table \ref{Tbl:Reacs}. As discussed below, we assume all free electrons come from hydrogen so that $n_e = n_p$.

The heating rate is dominated by photoelectrons from the ionization of hydrogen. In the high-UV case we consider here, cooling will be dominated by Ly$\alpha$ emission \citep{2009ApJ...693...23M}.  
The temperature is then determined by balancing photoelectric heating ($Q_{\rm 1s}$)  and Ly$\alpha$ cooling ($\Lambda_{\rm Ly\alpha}$),

\begin{equation}
Q_{\rm 1s}\left(N_{\rm 1s}\right)n_{\rm 1s} = \Lambda_{\rm Ly\alpha}\left(T\right) n_{\rm e}n_{\rm 1s}\,\, .
\label{Eqn:ThermalEq}
\end{equation}

\noindent where the Ly$\alpha$ cooling rate is $\Lambda_{\rm Ly\alpha}\left(T\right) \simeq 10.2\,{\rm eV}\left(C_{\rm 1s\rightarrow 2s}(T)+C_{\rm 1s\rightarrow 2p}(T)\right)$ and $C_{\rm 1s\rightarrow 2s}$ and $C_{\rm 1s\rightarrow 2p}$ are the collisional excitation rates from the 1s state to the 2s and 2p states, respectively (see table \ref{Tbl:Reacs}). Excitation to the 2s state also contributes to Ly$\alpha$ cooling since the $\ell$-mixing reactions rapidly turn 2s into 2p, which then emits a Ly$\alpha$ photon (see Figures \ref{Fig:reacrates2s}
and \ref{Fig:reacrates2p}).
In eq.\ref{Eqn:ThermalEq}, $Q_{\rm 1s}$ is the heating rate per particle, defined in eq.\ref{Eqn:Heat} below.

In cases where there is a non-zero Ly$\alpha$ excitation rate $J_{\rm Ly\alpha}$ due to the stellar radiation field, there exists the possibility, for sufficiently low gas temperatures, of Ly$\alpha$ heating when the excitation temperature of the radiation field approaches or exceeds the gas temperature.   In the appendix, we estimate the contribution of true Ly$\alpha$ absorption to the overall heating rate as a function of $J_{\rm Ly\alpha}$ and T.  We find that for the solar value of $J_{\rm Ly\alpha}$, the cooling rate is not significantly altered by the inclusion Ly$\alpha$ heating for the bulk of the layer that is responsible for the H$\alpha$ absorption; however, the heating could be important at lower temperatures near the base of the atomic layer (see fig. \ref{Fig:coolingstudy}). To address this issue fully it is necessary to have a detailed treatment of Ly$\alpha$ radiation transfer which is beyond the scope of this paper.


While some parameters of the planet are well known (see table \ref{table:189733b}), such as planetary mass ($M_p$), radius ($R_p$) and semi-major axis ($a$), other parameters are less certain, and may even vary with time, such as the ionizing and Ly$\alpha$ flux of the star HD 189733. Also uncertain is the role of day-night transport of heat in the upper atmosphere \citep{2007ApJ...661..515K}. Given these uncertainties, we parametrize the ionizing flux of HD 189733, and hence the heating and ionization rates, in an attempt to explain the observed transit depth and line profile. Specifically, the ionization rates will be multiplied by a factor $\xi/4$.  The flux averaged over $4\pi$ steradians corresponds to $\xi=1$. The substellar flux corresponds to $\xi=4$. Ionization rates larger than the nominal substellar value have $\xi \geq 4$.  For the Ly$\alpha$ rate, we use the solar spectrum and evaluate the flux at the orbital separation of HD 189733b.  We take the value of $J_{\rm Ly\alpha}$ to be constant throughout the atmosphere.

\begin{figure}
\epsscale{1.0}
\plotone{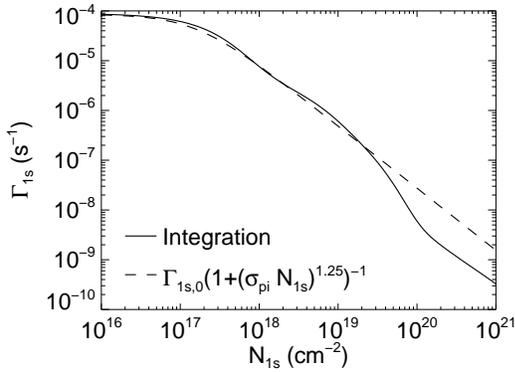}
\caption{1s state photoionization rate versus $N_{\rm 1s}$ for HD 189733b, as calculated from Eqn. \ref{Eqn:Ionize}.
The solid line is the numerical calculation, and the dashed line is a fit to the data. The threshold cross
section used in the fit is $\sigma_{\rm pi} = 6.3\times 10^{-18}\,{\rm cm^{-2}}$.}
\label{Fig:GammaRates}
\end{figure}

\begin{figure}
\epsscale{1.0}
\plotone{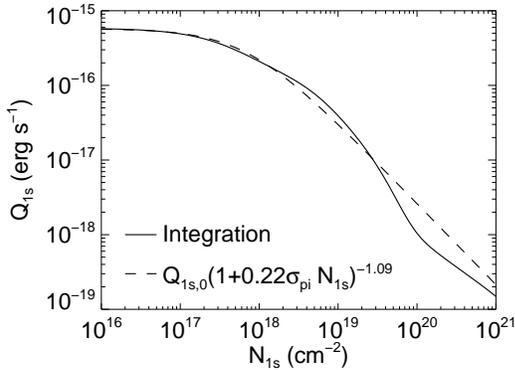}
\caption{1s photoelectric heating rate versus $N_{\rm 1s}$ for HD 189733b. The solid line is the numerical calculation using Eqn. \ref{Eqn:Heat} and
the dashed line is a fit. The threshold cross section used in the fit is $\sigma_{\rm pi} = 6.3\times 10^{-18}\,{\rm cm^{-2}}$.}
\label{Fig:QRates}
\end{figure}

The ground-state photoionization and photoelectric heating rates are defined as

\begin{equation}
\label{Eqn:Ionize}
\Gamma_{\rm 1s}\left(N_{\rm1s}\right) = \frac{\xi}{4}\int_{\nu_0}^{\infty} \frac{4\pi J_{\nu}}{h\nu}\sigma_{\rm 1s} e^{-\sigma_{\rm 1s}N_{\rm 1s}}d\nu 
\end{equation}
\noindent and 
\begin{equation}
\label{Eqn:Heat}
Q_{\rm 1s}\left(N_{\rm 1s}\right) = \frac{\xi}{4}\int_{\nu_0}^{\infty}\frac{4\pi J_{\nu}}{h\nu} h\left( \nu - \nu_0 \right)
\sigma_{\rm 1s}e^{-\sigma_{\rm 1s}N_{\rm 1s}} d\nu\,\, 
\end{equation}
where $J_\nu$ is the mean stellar intensity, $N_{\rm 1s} ( r ) = \int_r^\infty dr' n_{\rm 1s}(r')$ is the 1s state column density, $\nu_0 = 13.6\,{\rm eV}/h$ is the ionization threshold frequency, and $\sigma_{\rm 1s}$ is the photoionization
cross sections for the 1s state \citep{2006agna.book.....O}.
The ultraviolet spectrum is a synthetic spectra for HD 189733b downloaded from the X-exoplanets Archive at the CAB \citep{2011A&A...532A...6S}. The integration over frequency allows higher energy photons to dominate the integral at larger depth \citep{2011ApJ...728..152T}, leading to a larger rate
than if a single frequency was used \citep{2009ApJ...693...23M}.
Figs. \ref{Fig:GammaRates} and \ref{Fig:QRates} show the photoionization and heating rates due to the 1s state as a function of $N_{\rm 1s}$, evaluated  for HD189733b.  At small $N_{\rm 1s}$, the optically thin limit, the rates are constant, while at large columns they decrease roughly as a power law. 

Although ionization of the $n=2$ state hydrogen contributes negligibly to the electron density, ionization has an important effect on the number densities $n_{\rm 2s}$ and $n_{\rm 2p}$ of the 2s and 2p states. The rates are given by

\begin{equation}
\Gamma_{\rm 2s,2p} = \int_{\nu_1}^{\nu_0}\frac{4\pi J_\nu}{h\nu} \sigma_{\rm 2s,2p}\left(\nu\right) d\nu
\label{Eqn:Ionize2s2p}
\end{equation}

\noindent where $\nu_1=10.2\,{\rm eV}/h$ is the frequency threshold for $n=2$ state ionization.  The upper frequency bound of $\nu_0$ is included since photoionization of the ground state will both quickly attenuate ionizing radiation above this frequency and will contribute negligibly to the $n=2$ photoionization rate due to the small cross-section.  The spectrum used is taken from the Castelli and Kurucz Atlas \citep{2003IAUS..210P.A20C}.  We omit the factor of one quarter found in Eqn. \ref{Eqn:Ionize} due to the fact that Balmer continuum photons are optically thin throughout the region of interest, making the photoionization rate insensitive to the specific geometry considered.   Additionally, we have ignored attenuation of the radiation field due to the small column densities involved.

\subsection{ Level Populations}
\label{sec:levelpop}

In the upper atmosphere, several different physical processes  are important in setting the level populations, including collisional excitation and de-excitation, and bound-bound and bound-free transitions due to the stellar radiation field. The 2s and 2p states must be considered separately due to the lack of a fast radiative transition between 2s and the ground state\footnote{The two photon transition has a rate of  $A_{\rm 2s \rightarrow 1s}=8.26 {\rm \, s^{-1}}$, which is much smaller than $A_{\rm 2p \rightarrow 1s} = 6.3\times 10^8 {\rm \, s^{-1}}$.}.
The equation expressing rate equilibrium for the 2p state is

\begin{eqnarray}
&& B_{\rm 1s \rightarrow 2p} J_{\rm Ly\alpha} n_{\rm 1s} + C_{\rm 1s\rightarrow 2p}n_{\rm 1s}n_{\rm e} 
\notag\\ && 
+   C_{\rm 2s \rightarrow 2p} n_{\rm 2s} n_{\rm e}  + 
\alpha_{\rm 2p}n_{\rm e}^2 
\notag\\ & =& 
A_{\rm 2p\rightarrow 1s}n_{\rm 2p}
+ B_{\rm 2p \rightarrow 1s} J_{\rm Ly\alpha} n_{\rm 2p}
 \notag\\ & & 
+ C_{\rm 2p\rightarrow 1s}{n_{\rm 2p}}  n_{\rm e} 
+ C_{\rm 2p \rightarrow 2s} n_{\rm 2p} n_{\rm e}
 + \Gamma_{\rm 2p} n_{\rm 2p}  
 \label{eq:2p}
\end{eqnarray}

while that for the 2s state is

\begin{eqnarray}
&& C_{\rm 1s\rightarrow 2s}n_{\rm 1s}n_{\rm e} + C_{\rm 2p\rightarrow 2s}n_{\rm 2p}n_{\rm e} 
+ \alpha_{\rm 2s}n_{\rm e}^2 
\notag \\ & =& 
C_{\rm 2s\rightarrow 1s}n_{\rm e}{n_{\rm 2s}}  
+ C_{\rm 2s\rightarrow 2p}n_{\rm e}{n_{\rm 2s}} 
+ \Gamma_{\rm 2s} n_{\rm 2s} \notag \\ & &
+ A_{\rm 2s\rightarrow 1s}n_{\rm 2s}\,\, .
\label{eq:2s}
\end{eqnarray}

Here $C_{\rm i \rightarrow j}$ denotes the collisional transition from state $i$ to state $j$, and both the $\rm 1s 
\leftrightharpoons 2s$,  $\rm 1s \leftrightharpoons 2p$ and  the $\ell$-changing reaction $\rm 2s \leftrightharpoons 2p$
are included. Collisional transitions induced by 
atomic and molecular hydrogen, and helium, are negligible. 
The terms $\alpha_{\rm 2p} n_{\rm e}^2$ and $\alpha_{\rm 2s} n_{\rm e}^2$ represent the effective recombination coefficients to that level, and include radiative cascade from higher levels \citep{2011piim.book.....D}. We ignore
ionization of helium, and assume the electrons contributed by other elements, such as sodium and potassium, are negligible in comparison to that from hydrogen, hence $n_{\rm e}=n_{\rm p}$. The $A$'s and $B$'s are the Einstein coefficients for radiative transitions, and $J_{\rm Ly\alpha}$
is $\rm 1s \rightarrow 2p$ transition rate associated with Ly$\alpha$ radiation.  
The rate coefficients are listed in table \ref{Tbl:Reacs}.
 
 Due to the negligible contribution from collisions, the 2p occupation is set by radiative excitation and de-excitation,

\begin{eqnarray}
\frac{n_{\rm 2p}}{n_{\rm 1s}} & \approx & \frac{B_{\rm 1s \rightarrow 2p}J_{\rm Ly\alpha}}
{A_{\rm 2p \rightarrow 1s}} 
\nonumber \\ & \simeq & 
10^{-9}\ \left( \frac{5\ R_\star}{a} \right)^2 e^{16.9-\left(10.2\ {\rm eV}/k_b T_{\rm Ly\alpha, \star}\right) }.
\label{eq:n2p}
\end{eqnarray}
For the analytic estimate of $J_{\rm Ly\alpha}$,
we have included a dilution factor $(R_\star/2a)^2 \sim 10^{-2}$, for stellar radius $R_\star$
and orbital separation $a$, and the line intensity at the stellar surface is $\simeq (2h\nu^3/c^2) 
\exp\left( -10.2\ {\rm eV}/k_b T_{\rm Ly\alpha, \star} \right)$, where $T_{\rm Ly\alpha, \star} \simeq 7000\ {\rm K}$
is the approximate excitation temperature for the solar Ly$\alpha$. The 2p density is small
mainly due to the small stellar excitation temperature, and additionally because of the dilution factor. We show the profile of $n_{\rm 2p}$ in figure \ref{Fig:profilefiducial} including all the physical effects in eq.\ref{eq:2p}, verifying that radiative rates dominate, and that the 2p population is far smaller than 2s.

At sufficiently large $n_{\rm e}$, the 2s state is primarily populated by collisional excitation from the ground state while de-population is primarily due to the $\ell$-mixing transition 2s$\rightarrow$2p.  The 2s abundance can be estimated as

\begin{eqnarray}
\frac{n_{\rm 2s}}{n_{\rm 1s}}  & = & \frac{C_{\rm 1s\rightarrow 2s}}{C_{\rm 2s\rightarrow 2p}} = \frac{C_{\rm 2s\rightarrow 1s}}{C_{\rm 2s\rightarrow 2p}}\exp\left(-\frac{118400\,{\rm K}}{T}\right) \\
 & = & 1.627\times 10^{-8}\left(\frac{T}{10^4\,{\rm K}}\right)^{0.045}e^{11.84-118400\, {\rm K}/T} \notag\\
 & & \times \frac{8.633}{\log\left(T/T_0\right)-\gamma}\,\, 
\end{eqnarray}
\noindent where $T_0=1.02\,{\rm K}$ and $\gamma=0.57721...$ is the Euler-Mascheroni constant.

The strong temperature dependence in the exponential can be eliminated by using eq. \ref{Eqn:IonizeRecomb} and \ref{Eqn:ThermalEq},

\begin{eqnarray}
n_{\rm 2s}  & = &  \frac{C_{\rm 2s\rightarrow 1s}Q_{\rm 1s} n_{\rm 1s}}{C_{\rm 2s\rightarrow 2p}\Lambda_{\rm Ly\alpha, 0}\left(T\right)n_{\rm e}} \\
 & \approx & \frac{C_{\rm 2s\rightarrow 1s}\alpha_{\rm 1s}^{1/2}(T) n_{\rm 1s}^{1/2}}{C_{\rm 2s\rightarrow 2p}\Lambda_{\rm Ly\alpha, 0}\left(T\right)}\frac{Q_{\rm 1s}}{\Gamma_{\rm 1s}^{1/2}} \label{Eqn:n2sscale}\,\, ,
\end{eqnarray}

\noindent where we have defined $\Lambda_{\rm Ly\alpha}\left(T\right) \equiv \Lambda_{\rm Ly\alpha,0}\left(T\right)\exp\left(-118400\,{\rm K}/T\right)$. The heating rate scales with $N_{\rm 1s}$ as  $Q\left(N_{\rm 1s}\right)\approx Q_0/(0.22\sigma_{\rm pi}N_{\rm 1s})^{k_1}$ and the photoionization rate scales as $\Gamma_{\rm 1s} \approx \Gamma_{\rm 1s,0}/\left(\sigma_{\rm pi}N_{\rm 1s}\right)^{k_2}$ (see figs. \ref{Fig:GammaRates} and \ref{Fig:QRates} and \citet{2011ApJ...728..152T}).  Along a radial path the column density can be approximated by $N_{\rm 1s}\approx H(T)n_{\rm 1s}$, the 2s volume density scales as $n_{\rm 2s} \propto n_{\rm 1s}^{(1+k_2)/2-k_1`}$.  The parameters $k_1$ and $k_2$ depend on the EUV spectrum,  $k_1=1.25$ and $k_2=1.09$ for the synthetic HD189733 spectrum giving $n_{\rm 2s} \propto n_{\rm 1s}^{0.035}$.\footnote{For the solar spectrum \citet{2011ApJ...728..152T} found $k_1=1.2$ and $k_2 = 1.5$.} As a result, we find that $n_{\rm 2s}$ is fairly constant over a large range of pressures in the atomic layer.

\subsection{ Numerical Method}
\label{sec:numerics}

We consider a one-dimensional hydrostatic profile, with uniform spherical irradiation\footnote{Since there is minimal attenuation of $\Gamma_{\rm 2s}$ and $\Gamma_{\rm 2p}$ due to the small columns of $n=2$ hydrogen involved, the assumption of spherical irradiation is equivalent to irradiation in the slant geometry, as discussed in \S \ref{sec:iontherm}.} from the outside. We integrate the equation of hydrostatic balance, and the definition of the column,

\begin{equation}
\frac{\partial P}{\partial r}  =  - \frac{G M_p \rho}{r^2} 
\label{Eqn:Hydrostatic}
\end{equation}
\begin{equation}
\frac{\partial N_{\rm1s}}{\partial r}  =  -n_{\rm 1s}
\label{Eqn:ColDens}
\end{equation}

\noindent where the pressure is given by $P = \left(\left(1+f_{\rm He}\right)n_{\rm 1s} + \left(2+f_{\rm He}\right)n_{\rm e} \right)k_{\rm B} T$ and the density is $\rho = m_{\rm H}n_{\rm H} + m_{\rm e}n_{\rm e} + m_{\rm p}n_{\rm p} + f_{\rm He}m_{\rm He}\left(n_{\rm H} + n_{\rm p}\right)$.  $f_{\rm He}=0.1$ is the fraction, by number, of helium per hydrogen nucleus, and although it is considered non-reactive in our model, it affects the profile through its contribution to the density and pressure.  The relative helium abundance is considered to be constant throughout the atmosphere and is taken to have the solar value \citep{2006agna.book.....O}.

To generate the atmospheric profile, the pressure is integrated from the outer boundary using Eqn. \ref{Eqn:Hydrostatic} through a fourth-order Runge-Kutta scheme.  Given the pressure, temperature and ionization state can be solved for using Eqns. \ref{Eqn:IonizeRecomb} and \ref{Eqn:ThermalEq}.   The column density is then updated using Eqn. \ref{Eqn:ColDens} and the process repeated until the base radius $R_{\rm p}$ is reached.  To achieve the desired pressure $P_{\rm b} = 1\,{\rm \mu bar}$ at the base, the pressure at the outer boundary is varied using the secant method and new atmospheric profiles are generated until the target base pressure is reached.

Once the final profile is determined, the profiles for $n_{\rm 2s}$ and $n_{\rm 2p}$ can be calculated for the entire atmosphere using Eqns. \ref{eq:2p} and \ref{eq:2s}.

With the radial temperature and the $n_{\rm 2s}$ profile now calculated, the ${\rm 2s}$ state column density in the slant geometry, $N_{\rm 2s}$,  and the optical depth, $\tau_\nu$, can be calculated as a function of impact parameter b,

\begin{equation}
N_{\rm 2}\left(b\right) = \int_{-\infty}^{^\infty} n_{\rm 2s}\left(r\right)d\ell
\end{equation}

\begin{equation}
\tau_\nu\left(b\right) = \int_{-\infty}^{\infty} n_{\rm 2s}\left(r\right)\sigma_{{\rm H\alpha},\nu}\left(r\right)d\ell\,\, ,
\end{equation}

\noindent where $\ell$ is the line-of-sight coordinate and is related to the radial coordinate and impact parameter $b$ by $r=\sqrt{b^2+\ell^2}$.  We do not include $n_{\rm 2p}$ in the integrals due to its negligible contribution to the overall $n=2$ abundance for HD189733b. The proportional reduction in flux, ignoring the opaque disk at $r < R_{\rm p}$,  is then given by

\begin{equation}
\frac{F^{\rm (out)}_{\nu}-F^{\rm (in)}_\nu}{F^{\rm (out)}_\nu} = \frac{2}{R_\star^2}\int_{R_{\rm p}}^{R_\star} \left(1-e^{-\tau_\nu(b)}\right)b{\rm d}b
\label{Eqn:dFoverF}
\end{equation}

\noindent Integrating eq. \ref{Eqn:dFoverF} with respect to wavelength recovers the equivalent width in eq. \ref{Eqn:width}.  We do not use this definition in practice, instead we choose to work with the absorption measure $M_{\rm abs}$,



\begin{eqnarray}
M_{\rm abs} & = & \frac{\int e^{-\tau_{\rm ISM}} \left(F^{\rm (out)}_\lambda-F^{\rm (in)}_\lambda\right)d\lambda}{\int e^{-\tau_{\rm ISM}}F^{\rm (out)}_\lambda d\lambda} \\
 & = & \frac{\int d\lambda \int_{R_{\rm p}}^{R_\star} e^{-\tau_{\rm ISM}} \left(1-e^{-\tau_\lambda(b)}\right)I^\star_\lambda b\,db}{\int d\lambda\int_{R_{\rm p}}^{R_\star}e^{-\tau_{\rm ISM}}I^\star_\lambda bdb}\,\, ,
\label{eq:Mabs}
\end{eqnarray}

\noindent where $\tau_{\rm ISM}$ is the optical depth of the ISM and $I^\star_\lambda$ is the unabsorbed stellar intensity.   For Ly$\alpha$, we use a Voigt profile with a temperature of $8000\, {\rm K}$ and assume an interstellar hydrogen column density $N_{\rm H} = 10^{18.3}\,{\rm cm^{-2}}$ \citep{2005ApJS..159..118W,2010A&A...514A..72L}. We assume no attenuation of H$\alpha$ due to the interstellar medium.   For a constant  $I^\star_\lambda$, eq.\ref{eq:Mabs} is equivalent to $M_{\rm abs} = W_\lambda \Delta\lambda$.   In calculating $M_{\rm abs}$ for Ly$\alpha$, we assume that $I^\star_\lambda$ is proportional to


\begin{equation}
I^\star_\lambda \propto \exp\left(-\frac{1}{2}\left(\frac{\Delta v}{64\,{\rm km\, s^{-1}}}\right)^2\right)\frac{c}{\lambda^2}
\end{equation}
\noindent from \citet{2010A&A...514A..72L}. For H$\alpha$ transits we assume that $I^\star_\lambda$ is constant with $\lambda$.

\section{Results}
\label{sec:results}

\begin{deluxetable}{lcc}
\tablecolumns{2}
\tablewidth{0pc}
\tablecaption{Parameters for HD189733b}
\tablehead{\colhead{$\xi^a$}  &   $W_\lambda\,({\rm \AA})$ & $M_{\rm abs}$}
\startdata
 $1$ &   $3.24\times 10^{-3}$ & $2.2\times 10^{-4}$ \\
$14$ &  $1.41\times 10^{-2}$ & $8.8\times 10^{-4}$ 
\enddata
\tablenotetext{a}{See eqs. \ref{Eqn:Ionize} and \ref{Eqn:Heat} for the definition of $\xi$.}
\end{deluxetable}

To study the effect of varied heating and ionization rates, we vary $\xi$ between 1 and 20.  For each model atmosphere, we calculate $M_{\rm abs}$ for both H$\alpha$ and Ly$\alpha$.  For each $\xi$, we consider heating and ionization rates as defined by eqs. \ref{Eqn:Ionize} and \ref{Eqn:Heat}, base radius $R_{\rm b}=R_{\rm p} + 0.054R_{\rm p}$,  (see eq. \ref{Eqn:RadIncrease}) and base pressure $P_{\rm b} = 1\, {\rm \mu bar}$.  

\begin{figure}
\epsscale{1.0}
\plotone{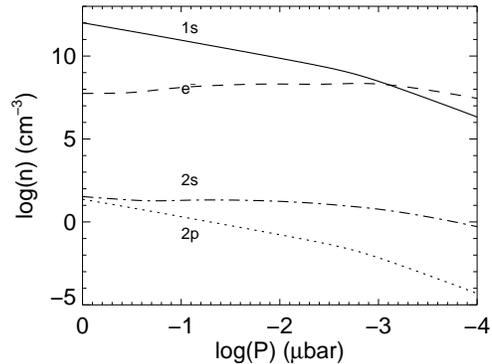}
\caption{ Number densities versus pressure for the $\xi=1$ case. Shown are the densities for 1s  (solid line), electrons (dashed line), 2s  (dash-dot line), and 2p (dotted line). }
\label{Fig:profilefiducial}
\end{figure}

\begin{figure}
\epsscale{1.0}
\plotone{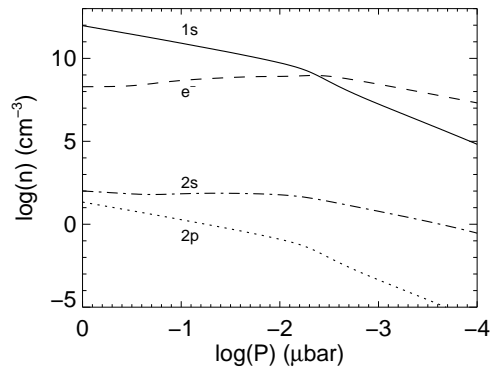}
\caption{ Number densities versus pressure for the $\xi=14$ case. Shown are the densities for 1s (solid line), electrons (dashed line), 2s  (dash-dot line), and 2p  (dotted line).}
\label{Fig:profilebestfit}
\end{figure}

Figure \ref{Fig:profilefiducial} shows the number density profile for the $\xi=1$ case.   Throughout both the atomic and ionized layers $n_{\rm 2s}$ is seen to roughly trace $n_{\rm e}$ while $n_{\rm 2p}$ follows the neutral hydrogen abundance, as expected from eq. \ref{eq:n2p}.  The $\xi=14$ profiles, shown in Figure \ref{Fig:profilebestfit}, exhibit the same behavior.   Within the atomic layer, $n_{\rm 2s}$ maintains a roughly constant abundance despite $n_{\rm 1s}$ varying by three orders of magnitude, although it is not explicitly constant, as can be seen in figs. \ref{Fig:n2linearfid} and \ref{Fig:n2linear}.  The 50\% ionization point is seen to move inward in pressure approximately linearly in with the increase in ionization rate (see figs. \ref{Fig:profilefiducial} and \ref{Fig:profilebestfit}).

\begin{figure}
\epsscale{1.0}
\plotone{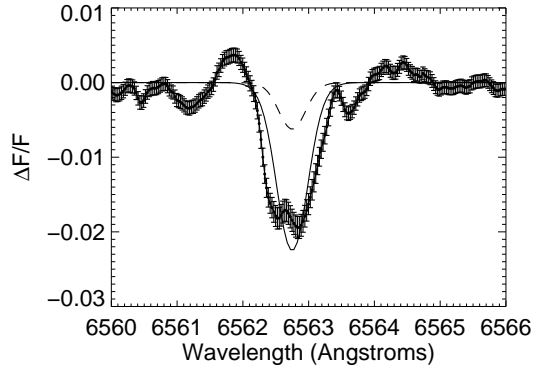}
\caption{ $\Delta F_\lambda/F_\lambda$  for the H$\alpha$ of HD189733b for $\xi=1$ (dashed line) and the $\xi=14$ (solid line) cases.  The observed data (filled circles) from \cite{2012ApJ...751...86J} are over plotted with error bars. }
\label{Fig:transitfid}
\end{figure}

The simulated transits for the abundance profiles in Figures \ref{Fig:profilefiducial} and \ref{Fig:profilebestfit} are shown in Figure \ref{Fig:transitfid}, compared to the data for HD 189733b from \citet{2012ApJ...751...86J}. The model profiles have been computed using eq. \ref{Eqn:dFoverF}. It is clear that $\xi=1$ underestimates the transit depth, while the higher ionization rate $\xi=14$ curve has roughly the correct width and depth. We note that there are significant oscillatory features in the measured data, which are much larger than the error bars shown in our Figure \ref{Fig:transitfid} and their Figure 3. These features persist in the wavelength regions outside the line. \citet{2012ApJ...751...86J} attribute these features to systematic errors not included in the error bars shown.

\begin{figure}
\epsscale{1.0}
\plotone{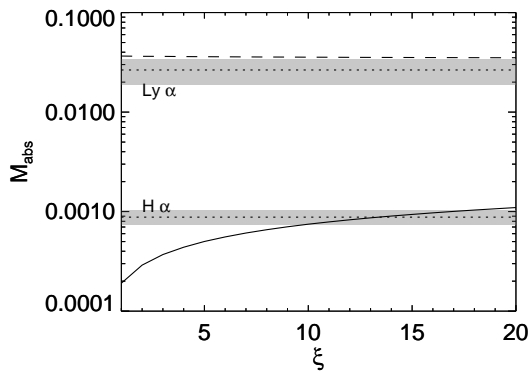}
\caption{$M_{\rm abs}$ versus the UV scaling parameter $\xi$ for HD 189733b.   The H$\alpha$ absorption (solid line) and the Ly$\alpha$ absorption (dashed line) are shown together with the observed values (dotted lines) labeled.  The grey bands indicate the error bars for the observed results. $M_{\rm abs}$ agrees with observation for $\xi = 14$.  Although the values for Ly$\alpha$ do not agree at $\xi=14$, the calculated value is within the margin of error for the observed quantity.  }
\label{Fig:ParameterStudy}
\end{figure}

Next we vary $\xi$ in order to derive the best fit to the corrected width $M_{\rm abs} = 8.8\times 10^{-4}$ reported by \citet{2012ApJ...751...86J},  with the corresponding values of $M_{\rm abs}$ shown in Fig. \ref{Fig:ParameterStudy}.  Additionally, we calculate $M_{\rm abs}$ for the unresolved Ly$\alpha$.  We find the H$\alpha$ data are best fit by $\xi=14$.   
\citet{2010A&A...514A..72L} determined the unresolved Ly$\alpha$ transit depth to be $5.05\pm 0.75\%$, including a $2.4\%$ contribution from the photospheric disk.    For the $\xi=14$ model, we find $M_{\rm abs} = 5.94\%$, including the opaque disk contribution, within two standard deviations of the observed value.  

\begin{figure}
\epsscale{1.0}
\plotone{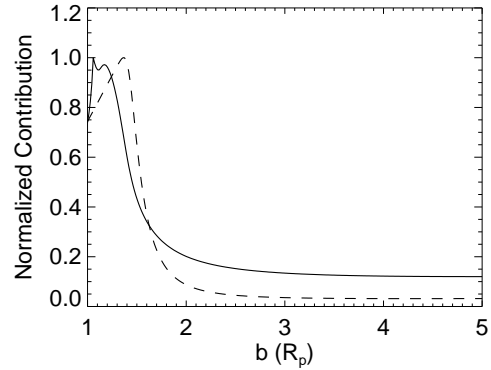}
\caption{The radial contribution to eq. \ref{eq:Mabs} as a function of b for the $\xi=14$ case for both H$\alpha$ (solid line) and Ly$\alpha$ (dashed line).  The plot is normalized so that the largest contribution has a value of 1. }
\label{Fig:RadCont}
\end{figure}

It is of interest to compare the location of the regions which contribute to $M_{\rm abs}$  for the H$\alpha$ and Ly$\alpha$. 
Figure \ref{Fig:RadCont} shows the contribution to the integral in eq. \ref{eq:Mabs} as a function of $b$ for both H$\alpha$ and Ly$\alpha$.  The primary contribution for Ly$\alpha$ occurs farther out than that of H$\alpha$ since Ly$\alpha$ becomes optically thick throughout the integrated band at lower column densities than H$\alpha$.  The linear decrease for decreasing $b$ is a geometric effect arising from the fact that annuli at larger impact parameters contribute more.  For H$\alpha$, the largest contribution occurs at the base of the atomic layer with a noticeable decrease for impact parameters inside that layer.   For large b, the contributions for both H$\alpha$ and Ly$\alpha$ are small.  The constancy at large $b$ is due to the large scale height found at these radii.   If a planetary wind is present, the densities at these radii could be much lower than found in our hydrostatic model (e.g., \citealt{2004Icar..170..167Y}), further decreasing their contribution to $M_{\rm abs}$.  




\begin{figure}
\epsscale{1.0}
\plotone{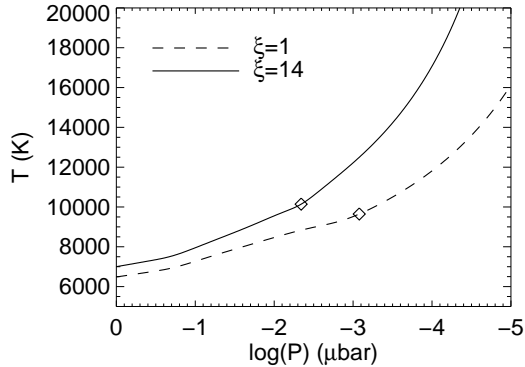}
\caption{The temperature profile versus pressure for $\xi=1$ and $\xi=14$.   For each curve, the location where the atmosphere transitions from being dominated by ions to being dominated by atomic hydrogen is denoted by a diamond.}
\label{Fig:Temps}
\end{figure}

Although $\xi=14$ constitutes an order-of-magnitude increase in the heating rate,  the change in temperature in the atomic layer is relatively modest.  If we take the volume-weighted average temperature,

\begin{equation}
\left<T\right> = \frac{3}{R_{\rm top}^3-R_{\rm b}^3}\int_{R_{\rm b}}^{R_{\rm top}} T\left(r\right) r^2 dr
\end{equation}

\noindent where $R_{\rm top}$ is the radius where the electron density equals the neutral hydrogen density (see fig. \ref{Fig:cartoon}), $n_{\rm e} = n_{\rm H}$.  For the $\xi=1$ case, we find $\left<T\right> = 8637\, {\rm K}$, and for $\xi = 14$  we find $\left<T\right> =8902\,{\rm K}$.  Previous investigators have found the need for additional UV heating \citep{2010A&A...514A..72L}. Although we enforce this increase in temperature through an increase in the photoelectric heating rate for hydrogen, a more complete modeling of the heating and cooling processes in the atmosphere could explain the required temperature difference \citep{2012arXiv1210.1536K,2012arXiv1210.1543K}.

Fig.  \ref{Fig:Temps} shows the temperature profiles as a function of gas pressure for both $\xi=1$ and $\xi=14$.  Within the atomic layer ($n_{\rm 1s} > n_{\rm e}$), the $\xi = 14$ case is approximately $1000\, {\rm K}$ above the temperatures in the $\xi = 1$ case at comparable pressures; however, due to the proportionately smaller ionization rate, the atomic layer extends to lower pressures for $\xi=1$ which accounts for the similar values of $\left<T\right>$.

We note that the temperatures at $R_{\rm b}$ are $6000-7000\, {\rm K}$, much higher than the temperatures required for the formation of molecules.  This is a limitation of our model due to the simplified prescriptions for heating and cooling.

\begin{figure}
\epsscale{1.0}
\plotone{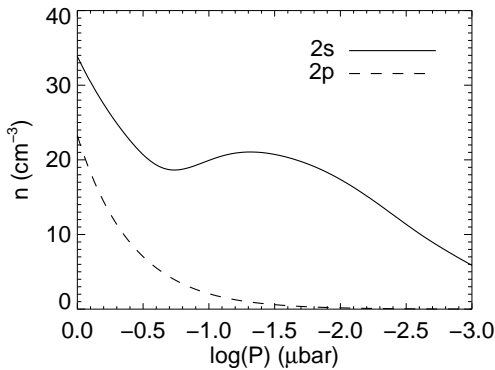}
\caption{The radial distribution of 2s (solid line) and 2p (dashed line) for the $\xi=1$ case. These are the same data as fig. \ref{Fig:profilefiducial} but with a linear scale to better show the the variation in abundance. }
\label{Fig:n2linearfid}
\end{figure}

\begin{figure}
\epsscale{1.0}
\plotone{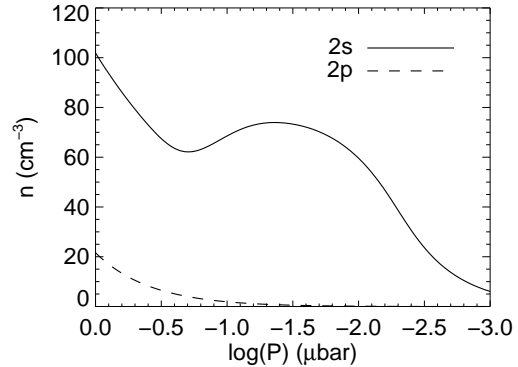}
\caption{The radial distribution of 2s (solid line) and 2p (dashed line) hydrogen for the $\xi = 14$ case.   These are the same data as fig. \ref{Fig:profilebestfit} but with a linear scale to better show the the variation in abundance. }
\label{Fig:n2linear}
\end{figure}

The abundance profiles are shown for $\xi=1$ and $\xi=14$ in figs. \ref{Fig:profilefiducial} and \ref{Fig:profilebestfit}, respectively.  In both cases, $n_{\rm 2p}$ traces the abundance of neutral hydrogen, as expected from eq. \ref{eq:n2p}.   Within the atomic layer, $n_{\rm 2s}$ maintains a roughly constant abundance despite $n_{\rm 1s}$ varying by three orders of magnitude. To better exhibit the small changes in $n_{\rm 1s}$, Figures \ref{Fig:n2linearfid} and \ref{Fig:n2linear} show a linear scale.

\begin{figure}
\epsscale{1.0}
\plotone{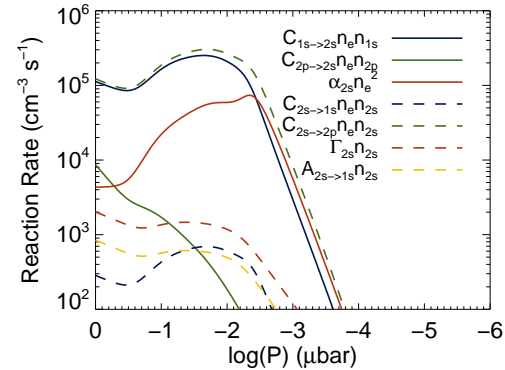}
\caption{ The reaction rates, for the 2s state, per unit volume versus pressure, using $\xi=14$.  Collisional excitation from the 1s state to the 2s state (solid blue line) is the dominant creation pathway and is balanced by the collisional transition from the 2s to the 2p state (green dashed line).   Additional creation pathways are the collisional transition from 2p to 2s (solid green line) and recombination to the 2s state (solid red line).  The remaining destruction pathways are photoionization (dashed red line),  collisional de-excitation from the 2s to the 1s state (dashed blue line), and the two-photon radiative transition to the 1s state (dotted blue line). }
\label{Fig:reacrates2s}
\end{figure}

\begin{figure}
\epsscale{1.0}
\plotone{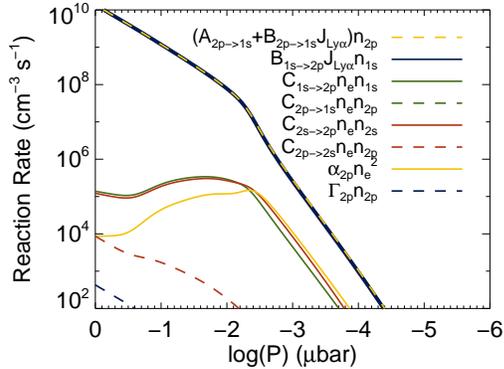}
\caption{ The reaction rates, for the 2p state, governing the creation and destruction of hydrogen in the 2p state for the best-fit case.  Radiative excitation from the 1s state (solid blue line) is the dominant creation pathway and radiative de-excitation (dashed blue line) is the dominant destruction mechanism.   In addition, 2p is created through collisional excitation from the 1s state (solid green line), the collisional transition from 2s to 2p (solid red line), and recombination to the 2p state (solid yellow line).  The remaining destruction mechanisms are collisional de-excitation to the 1s state (dashed green line), the collisional transition from 2p to 2s (dashed red line), and photoionization of the 2p state (dashed yellow line).}
\label{Fig:reacrates2p}
\end{figure}

Fig. \ref{Fig:reacrates2s} shows the reaction rates for all reactions involved in the formation of 2s hydrogen for the $\xi=14$ model.    Within the hydrogen layer, the abundance is set by the balance of collisional excitation from 1s and the $\ell$-mixing reaction.  As pressures decrease, and the atmosphere becomes ionized, the contribution of radiative recombination increases.  Once the atmosphere becomes predominantly ionized, radiative recombination becomes the dominant formation pathway.  

The reaction rates for 2p hydrogen are shown in fig. \ref{Fig:reacrates2p}.  Throughout the atmosphere,  $n_{\rm 2p}$ is set by the radiative transition between 1s and 2p, with little contribution from other rates.   Attenuation of the Ly$\alpha$ radiation could play a role in decreasing $n_{\rm 2p}$; however, at $\mu$bar pressures we find that the radiative transition rates exceed the collisional rates by a factor of $10^5$. 


\section{The Case of HD209458b}
\label{sec:hd209}

\begin{figure}
\epsscale{1.0}
\plotone{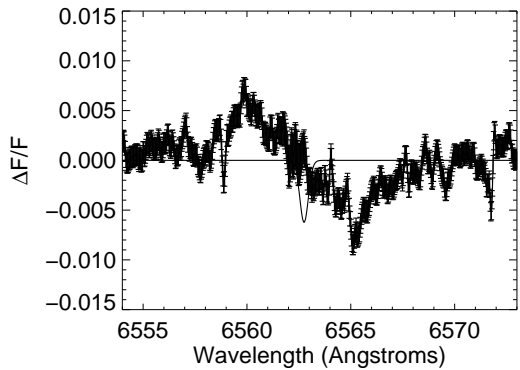}
\caption{ Model for $\Delta F_\lambda/F_\lambda$  for the H$\alpha$ of HD209458b for $\xi=1$.  The observed data (filled circles) from \cite{2012ApJ...751...86J} are over plotted with error bars.  Note that the range of wavelengths in the plot is larger than in fig. \ref{Fig:transitfid} in order to capture the absorption and emission features.  }
\label{Fig:transitfid209458}
\end{figure}

\begin{figure}
\epsscale{1.0}
\plotone{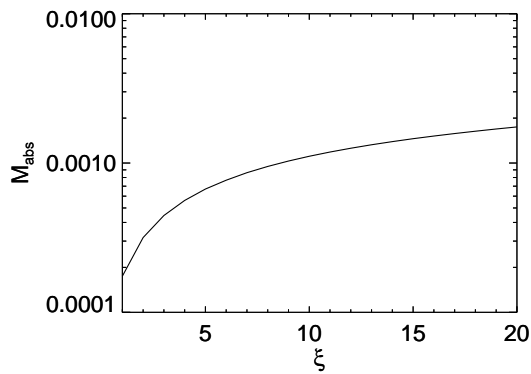}
\caption{$M_{\rm abs}$ for H$\alpha$ absorption versus the UV scaling parameter $\xi$ for HD209458b.  }
\label{Fig:ParameterStudy209458}
\end{figure}

A similar analysis has been performed for HD209458b. Since the line profiles for model and data are discrepant,  we only produce a parameter study of $M_{\rm abs}$ for the same range of $\xi$ used above.  

Using parameters taken from exoplanet.eu, we take $M_{\rm p} = 0.714\, M_{\rm J}$, $R_{\rm p} = 1.38\,R_{\rm J}$, 	$a = 0.04747\, {\rm au}$, and use the simulated UV spectrum for HD209458 from the X-exoplanets Archive at the CAB \citep{2011A&A...532A...6S}.

Fig. \ref{Fig:transitfid209458} shows the absorption profile for $\xi=1$.   It is immediately obvious that there is no match between the model prediction and the observers profile. The model profile is symmetric about line center while and the observed profile is antisymmetric. The observed profile has a width characteristic of the orbital velocity ($150\ {\rm km\ s^{-1}}$) while the model is only a couple Doppler widths ($10\ {\rm km\ s^{-1}})$ wide. Fig. \ref{Fig:ParameterStudy209458} shows the dependence of $M_{\rm abs}$ on $\xi$.  The curve is qualitatively similar to that of the HD189733b case.  

It should be noted that for HD209458b we find that 2p becomes more important than for HD189733b.  Accurately accounting for 2p requires modeling the radiative transfer of the Ly$\alpha$ which is beyond the scope of this paper.

\section{Balmer Continuum Absorption}
\label{sec:balmercont}
Given the observation of Balmer continuum absorption by HD209458b \citep{2007Natur.445..511B}, we estimate the transit depth for HD189733b with our model.   At the Balmer edge, the bound-free absorption cross-section is $\sigma_{\rm bf} \sim 10^{-17}\,{\rm cm^2}$ with the cross section decreasing for smaller wavelengths.   For $\xi=14$, the number density is bounded $n_2 \lesssim 10^2\,{\rm cm^{-3}}$ and a characteristic length is $L \sim 10^9-10^{10}\,{\rm cm}$.   This yields an upper limit on the optical depth $\tau_{\rm bf} \lesssim \sigma_{\rm bf}n_2L = 10^{-5} - 10^{-6}$ with smaller values of $\tau_{\rm bf}$ for both larger impact parameters and shorter wavelengths.  The reduction in flux due to bound-free absorption relative to H$\alpha$ absorption should be proportional to the ratio of cross sections,

\begin{eqnarray}
\left(\frac{\Delta F_\nu}{F_\nu}\right)_{\rm bf} & = & \frac{\sigma_{\rm bf}(\nu)}{\sigma_{H\alpha}(\nu)} \left(\frac{\Delta F_\nu}{F_\nu}\right)_{\rm H\alpha}\\
 & \approx & 10^{-4}  \left(\frac{\Delta F_\nu}{F_\nu}\right)_{\rm H\alpha} \approx 10^{-6}\,\, .
\end{eqnarray}
\noindent  Within the context of our model, the transit depth due to the Balmer bound-free continuum absorption is too small to be observed for both HD189733b and HD209458b.

\section{Summary and Discussion}
\label{sec:summary}


We have modeled the abundance of n=2 hydrogen in hydrostatic atmospheres. We find that the H$\alpha$ absorption can be explained by metastable 2s, similar to what is found in the interstellar medium (e.g., \citealt{1957IAUS....4...92T}).   The dominant mechanism for the creation of 2s hydrogen is collisional excitation from the 1s state where it subsequently collisionally transitions to the 2p state and is finally radiatively de-excited.   The 2s population dominates 2p throughout the atmosphere by two orders of magnitude for the parameters used in our study, although the specifics depend on the chosen value of $J_{\rm Ly\alpha}$.   We do not model the spatial variation in the intensity of Ly$\alpha$, instead choosing a constant value.   This assumption allows us to estimate an upper limit on the 2p abundance.  Since 2p remains negligible compared to 2s for our chosen $J_{\rm Ly\alpha}$, we can assume that it should similarly be negligible in the case where resonance scattering has depleted the available Ly$\alpha$ photons.   Unlike $n_{\rm 2p}$, $n_{\rm 2s}$ has limited dependence on the radiation field, instead depending strongly on the gas temperature through the exponential dependence found in $C_{\rm 1s\rightarrow 2s}$.  We find that the data are best fit by an atomic hydrogen layer approximately $500-1000\,{\rm K}$ hotter than our $\xi=1$ case, which corresponds to $\xi=14$.   Caution should be used in the physical interpretation of this value due to the simplified heating and cooling present in our model.   

Because of the strong dependence on temperature, H$\alpha$ should be considered a complementary probe to Ly$\alpha$ which is relatively insensitive to the gas temperature, probing instead all the atomic gas. 

This model differs from the calculations of \citet{2012ApJ...751...86J}.   Assuming that both H$\alpha$ and Ly$\alpha$ are optically thin, they used eq. \ref{Eqn:WidthColumnDens} to derive an excitation temperature of $T_{\rm exc} = 2.6\times 10^4\,{\rm K}$.  Arguing that the levels will approach a Boltzmann distribution deeper in the atmosphere and the excitation temperatures found are not compatible with the expected gas temperature at these densities, they conclude that the absorption must occur in the planetary wind.    For our $\xi = 14$ model, we find $n_1/n_2 \sim 10^{7}- 10^{10}$ which using eq. \ref{Eqn:Boltz} gives 

\begin{equation}
T_{\rm exc} = \frac{118400\,{\rm K}}{\log\left(\frac{4n_1}{n_2}\right)} \simeq 4800-7000\,{\rm K}\,\, .
\end{equation}

\noindent This value for $T_{\rm exc}$ is lower than the temperature in the neutral layer and is significantly lower than the value quoted by \citet{2012ApJ...751...86J} due to the significantly smaller $n=2$ abundance in our model. The former is due to the collisional excitation being balanced by the $\ell$-mixing reactions, not collisional de-excitation.  The latter is due to the overestimation of the gas temperature due to the assumption of optically thin Ly$\alpha$.

\citet{2012arXiv1206.5003T} have proposed that the H$\alpha$ absorption can be explained by the same mechanism they use to explain Ly$\alpha$ absorption, colliding planetary and stellar winds.  This mechanism allows for Ly$\alpha$ absorption $100\,{\rm km\, s^{-1}}$ from line center to be explained by the formation of hot ($T\sim 10^6\,{\rm K}$) neutral hydrogen generated through charge exchange with solar wind protons.  If H$\alpha$ is in fact probing the interface between the two winds, the Ly$\alpha$ and H$\alpha$ lines should have comparable widths; however, the observed width for HD189733b is  $\simeq 40\,{\rm km\,s^{-1}}$. \citet{2012arXiv1206.5003T} propose to explain this population of cool hydrogen in its $n=2$ state through cooling of the initially hot population formed through charge exchange. 

We find that observed absorption of H$\alpha$ can be explained by our model with the primary signal coming from the neutral atomic layer.  Within this layer, the abundance of $n=2$ hydrogen is roughly constant, even though the overall abundance of hydrogen is increased by three orders of magnitude from the top of the layer to the bottom because the density increase is offset by a decrease in temperature.  The transition to molecular hydrogen will lead to a downturn in the overall $n=2$ abundance.  As a result, there should not be significant contribution to H$\alpha$ absorption within this layer.

Although many simplifications have been made, we have included the relevant physics and reproduced the transit signal observed by \citet{2012ApJ...751...86J} for HD189733b.   There are, however, many avenues for improving the calculation. We do not include the cooling required to cause the transition to the molecular state.  This results in our model having too high a temperature at the base radius.  The inclusion of the necessary physics will eliminate the need for the intermediate boundary at $R_{\rm b}$.  

Our model also required a higher rate of heating than expected from UV models from the X-exoplanets Archive at the CAB \citep{2011A&A...532A...6S}.   The inclusion of heavier atomic species allows for far-UV absorption higher in the atmosphere \citep{2012arXiv1210.1536K,2012arXiv1210.1543K} which could reduce the need for the large heating rate.

Our model has assumed spherical symmetry. Due to the strong dependence of the $n=2$ abundances on the temperature, temperature variations between the day and night side could induce strong day-night variation in $n_{\rm 2s}$ and $n_{\rm 2p}$, which will be observable since the transit probes the day-night terminator. This can partially be negated by the redistribution of thermal energy by zonal winds; however, this provides more reason to model the signal in three-dimensions, not less.

\acknowledgements

We would like to thank Adam Jensen and Seth Redfield for useful discussions, and for kindly providing the data for HD 189733b and HD 209458b, as well as Joshua N. Winn, Remy Indebetouw, and Mark Whittle for helpful discussions regarding H$\alpha$ observations.  We also thank the referee for providing constructive comments and suggestions. The authors acknowledge support from NSF AST-0908079 and NASA Origins NNX10AH29G grants.

\bibliographystyle{apj}
\bibliography{halpha}

\appendix
\section{ Net Ly$\alpha$ Heating or Cooling Including Collisional Excitation and Thermalization of Stellar Photons }
\label{appendix}

In this section we investigate the effect of the radiation field on the Ly$\alpha$ cooling rate.   The cooling rate $\Lambda(T)$, as used in the text, ignores the possibility of heating due to collisional de-excitation.   Allowing for this possibility, the cooling rate takes the form

\begin{equation}
\Lambda\left(T,J_{\rm Ly\alpha}\right)n_{\rm e}n_{\rm 1s} = 10.2\,{\rm eV}\left(C_{\rm 1s\rightarrow 2s}n_{\rm 1s} + C_{\rm 1s\rightarrow 2p}n_{\rm 1s}-C_{\rm 2s\rightarrow 1s}n_{\rm 2s} - C_{\rm 2p\rightarrow 1s}n_{\rm 2p} \right)n_{\rm e}\,\, .
\label{eq:apxcooling}
\end{equation}

This cooling rate depends on radiative excitation and de-excitation as well as recombination and photoionization implicitly through their effect on the level populations, $n_{\rm 2s}$ and $n_{\rm 2p}$ (see eqs. \ref{eq:2p} and \ref{eq:2s}).   For sufficiently low temperature or large $J_{\rm Ly\alpha}$, the collisional de-excitation rates can become comparable to the excitation rates, reducing the net cooling rate and, in extreme cases, result in net heating.  

To this end, we solve eqs. \ref{eq:2s} and \ref{eq:2p} for $n_{\rm 2s}$ and $n_{\rm 2p}$ given fixed values of $n_{\rm e}$ and $T$.  In the absence of recombination, the solution for $\Lambda(T,J_{\rm Ly\alpha})$ is independent of $n_{\rm 1s}$, the dependence having been explicitly factored out in eq. \ref{eq:apxcooling}.  Even if recombination is included, it can safely be ignored so long as its contribution to the $2s$ and $2p$ abundance remains small.   For the 2s state, where collisional excitation is the dominant formation pathway, this results in the condition

\begin{equation}
\frac{n_{\rm e}}{n_{\rm 1s}} \ll \frac{C_{\rm 1s\rightarrow 2s}}{\alpha_{\rm 2s}} = 5.17\times 10^5\left(\frac{T}{10^4\, {\rm K}}\right)^{0.075}\exp\left(-\frac{118400\,{\rm K}}{T}\right)\,\,.
\end{equation}


\noindent For the 2p state, radiative excitation sets the occupation, resulting in the requirement,

\begin{equation}
n_{\rm e} \ll 5.19\times 10^{10}\,{\rm cm^{-3}} \left(\frac{T}{10^4\,{\rm K}}\right)^{0.32}\left(\frac{J_{\rm Ly\alpha}}{J_{\rm Ly\alpha,0}}\right)^{1/2}\left(\frac{n_{\rm 1s}}{10^{10}\,{\rm cm^{-3}}}\right)^{1/2}\,\,.
\end{equation}

\noindent For the model of HD189733b, both of these conditions are satisfied, as is evidenced by figs \ref{Fig:reacrates2s} and \ref{Fig:reacrates2p}.  In cases where the contribution of recombination is non-negligible, the effect would be to increase the overall abundance of 2s and 2p hydrogen, thus suppressing Ly$\alpha$ cooling.

We take the electron density to be $n_{\rm e} = 10^8\,{\rm cm^{-3}}$, the typical abundance found in the atomic layer in our models.   Although the solutions have dependence on $n_{\rm e}$ beyond that shown in eq. \ref{eq:apxcooling}, we have found that varying the abundance changes the results minimally.

\begin{figure}
\epsscale{1.0}
\plotone{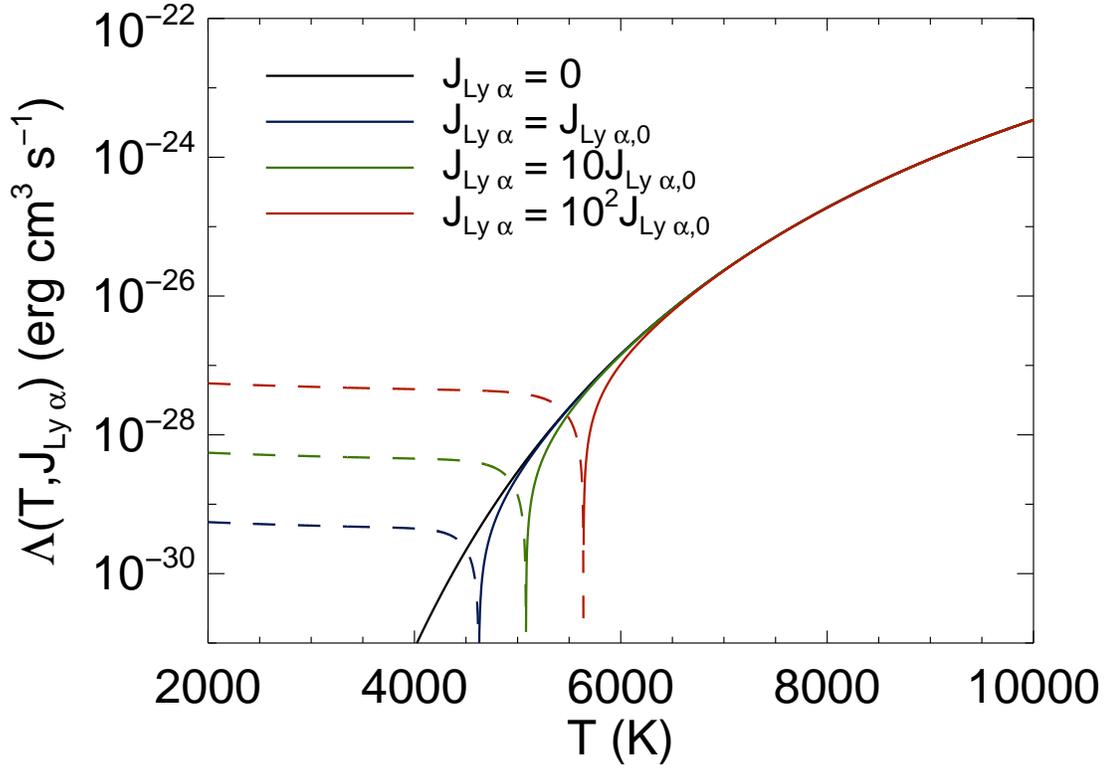}
\caption{The cooling rate as a function of gas temperature for differing values of $J_{\rm Ly\alpha}$.  The case of no ambient Ly$\alpha$ field,  $J_{\rm Ly\alpha} = 0$, is given in black.  The cases of $1$, $10$, and $10^2$ times $J_{\rm Ly \alpha,0}$ are shown in blue, green, and red, respectively.  For all curves, a solid line represents net cooling and a dashed line represents net heating.}
\label{Fig:coolingstudy}
\end{figure}

Fig. \ref{Fig:coolingstudy} shows the cooling function for four values of $J_{\rm Ly\alpha}$, with $J_{\rm Ly\alpha,0}=3.42\times 10^{-11}\,{\rm s^{-1}}$, the value used in our model above.  For the case of $J_{\rm Ly\alpha}=0$, we recover the standard cooling rate\footnote{Although we keep the collisional de-excitation terms in our cooling rate, even for the $J_{\rm Ly\alpha}=0$ case, it can be shown to be negligible in this case.}. For the three cases where $J_{\rm Ly\alpha}$ is non-zero, we find that the cooling rate diverges from the standard case as the temperature decreases with the cooling rate becoming zero as the gas temperature approaches the excitation temperature of the radiation field, $T_{\rm ex}  \sim 118400\,{\rm K}/\log\left((2h\nu^3/c^2)J_{\rm Ly\alpha}g_1/g_2\right)$.   For temperatures below this threshold,  collisional de-excitations return energy to the gas faster than it is removed through collisional excitation, resulting in net heating.   The weak dependence of the $C_{\rm 2s\rightarrow 1s}$ and $C_{\rm 2p\rightarrow 1s}$ results in the heating rate being roughly constant  as the temperature decreases.  

We note that for our model the temperature did not drop below $T=6000\,{\rm K}$ so the inclusion of Ly$\alpha$ heating would not have changed our results.  

The implications of this result, however, are that at low temperatures the Ly$\alpha$ cooling rate is possibly over-estimated and the existence of a transition to heating creates a temperature floor in the limit where Ly$\alpha$ is the only cooling mechanism.  The specific location of this temperature floor depend on the value of $J_{\rm Ly\alpha}$, potentially making the details of the radiative transfer problem important.  Ly$\alpha$ undergoes resonant scattering in the atmosphere, and due to the low absorption probability, is unlikely to attenuate exponentially, leaving the possibility that the intensity is non-negligible within the atomic hydrogen layer and could play a role as the transition to molecular hydrogen is approached.  

\label{lastpage}
\end{document}